\documentclass[a4paper,10pt,english]{article}
\usepackage[margin=1in]{geometry}
\usepackage[utf8]{inputenc}
\usepackage{enumitem}
\usepackage[colorlinks]{hyperref}
\usepackage{graphicx}
\usepackage{subcaption}
\usepackage{color}
\usepackage{amsmath}
\usepackage{authblk}
\usepackage{gensymb}
\usepackage{setspace}
\usepackage{amssymb}
\usepackage{longtable}
\usepackage{tabularx}
\usepackage{tablefootnote}
\usepackage[table,xcdraw]{xcolor}
\usepackage[
backend=biber,
style=numeric,
sorting=none
]{biblatex}
\usepackage{colortbl}
\usepackage{fancyhdr}
\newenvironment{keyword}
  {\par\noindent\textbf{Keywords: }\ignorespaces}
  {\par}

\title{Quantum Protocols for Time Synchronisation and Distribution:\\A Critical Assessment}
\author[1,2]{Michal Krelina\thanks{michal.krelina@qubitrium.tech}}
\author[1]{Utku Tefek}
\author[1,3]{Zeki C. Seskir}
\author[1,4]{Kadir Durak}
\affil[1]{QUBITRIUM B.V., Delft, Netherlands.}
\affil[2]{QuDef B.V.,  Delft, Netherlands.}
\affil[3]{Karlsruhe Institute of Technology, Institute for Technology Assessment and Systems Analysis, Karlsruhe, Germany.}
\affil[4]{Electrical and Electronics Engineering Department, Özyeğin University, Istanbul, Turkey.}

\begin{document}
\maketitle
\thispagestyle{fancy}

\begin{abstract}
Precise time synchronisation underpins critical infrastructure from telecommunications and financial markets to power grids and scientific metrology.
Several families of quantum protocols have been proposed and demonstrated for clock synchronisation and time distribution, exploiting entangled photon pairs, quantum key distribution (QKD) correlations, Hong-Ou-Mandel interference, and entangled clock networks.
We critically assess these approaches, reviewing the main quantum time synchronisation (QTS) protocol families, quantifying the gap between theory and experiment, and identifying practical bottlenecks in sources, detectors, and channels.
We survey the classical timing landscape from Network Time Protocol (NTP) and GPS to laboratory-grade optical frequency transfer, and compare quantum and classical methods at equivalent maturity.
We examine use cases including financial trading, power grids, telecommunications, scientific metrology, and military applications, evaluating whether quantum timing offers a realistic advantage.
We show that time transfer, not clock performance, is now the bottleneck for distributed optical timekeeping: the best demonstrated synchronisation uncertainty (2.46~ps) falls two to three orders of magnitude short of what optical clocks with fractional frequency uncertainties of $10^{-18}$--$10^{-19}$ require.
Our assessment is that quantum time synchronisation will not replace classical methods for most applications in the near-to-medium future.
Its near-term value lies in physical-layer security against timing manipulation and integration with quantum communication networks, while closing the synchronisation gap for scientific metrology remains the most critical open challenge.
\end{abstract}




\section{Introduction}\label{sec:intro}

Precise time synchronisation is a critical but often invisible requirement for modern infrastructure.
Telecommunications networks rely on sub-microsecond synchronisation to coordinate 5G base stations. 
Financial exchanges timestamp trades to microsecond accuracy under regulatory mandates such as the EU's MiFID~II.
Power grids use GPS-synchronised phasor measurement units to monitor voltage phase angles across continental-scale networks.
Scientific facilities, from particle accelerators to gravitational-wave observatories, require nanosecond-to-sub-picosecond timing over kilometres of fibre. Global navigation satellite systems (GNSS) themselves depend on nanosecond-level synchronisation between orbiting atomic clocks to derive position fixes.
In each of these domains, timing is currently provided by classical methods, primarily GPS and the Precision Time Protocol (PTP), with more specialised techniques such as White Rabbit~\cite{Lipinski2011} and coherent optical frequency transfer~\cite{Droste2013} serving high-precision applications.

These classical systems, however, share a common vulnerability: they offer no physical-layer protection against timing manipulation.
GPS signals are weak and can be jammed or spoofed with commercially available equipment~\cite{Humphreys2011}.
PTP and the Network Time Protocol (NTP) authenticate data content but not propagation delay, leaving them open to delay attacks~\cite{Lee2019security}.
The growing reliance of critical infrastructure on precise timing, combined with an increasing number of documented GPS interference incidents~\cite{IATA2025safety}, has motivated the search for timing solutions that provide both precision and inherent security.

Quantum protocols for time synchronisation have been developed over the past two decades and, in principle, offer two distinct advantages over classical methods.
First, protocols based on entangled photon pairs can achieve dispersion-immune timing through nonlocal quantum correlations~\cite{Giovannetti2001nature,Giovannetti2001prl}, an effect with no classical analogue.
Second, protocols derived from quantum key distribution (QKD) encode timing information in single photons whose interception is detectable through the laws of quantum mechanics, providing physical-layer security that does not rely on computational assumptions~\cite{LamasLinares2018}.

This paper provides a critical assessment of these quantum approaches: what they can deliver, what they cannot, and where they add value compared to classical alternatives.
We review the main protocol families (Section~\ref{sec:protocols}), analyse their practical limitations (Section~\ref{sec:challenges}), survey the classical timing landscape (Section~\ref{sec:classical-protocols}), provide a quantum-vs-classical comparison (Section~\ref{sec:comparison-qvc}), assess the widening gap between optical clock precision and time transfer capability (Section~\ref{sec:clocks}), and examine specific use cases (Section~\ref{sec:use-cases}).
A recent survey by Khalid et al.~\cite{Khalid2026} catalogues the growing quantum clock synchronisation protocol landscape in detail; our paper complements that work by providing a systematic comparison with classical methods and assessing the practical deployment prospects for specific applications.

Before proceeding, it is worth clarifying several terms from time and frequency metrology that appear throughout this paper.
\textit{Clock synchronisation} refers to the process of adjusting two or more clocks so that they indicate the same time, that is, reducing the \textit{clock offset}, which is the time difference between a local clock and a reference clock at a given instant.
\textit{Time transfer} (or time distribution) is the broader task of conveying a time reference from one location to another, typically over an optical fibre or a free-space link.
The quality of a time transfer system is characterised by two complementary metrics.
\textit{Accuracy} describes how close the transferred time is to the true value (i.e. how small the systematic offset is), while \textit{precision} (or \textit{stability}) describes how much the transferred time fluctuates from one measurement to the next.
The standard measure of timing stability is the \textit{time deviation} (TDEV), denoted $\sigma_x(\tau)$, which quantifies the timing fluctuations as a function of the averaging time $\tau$, longer averaging generally yields lower TDEV and thus better stability.
A related quantity used for frequency comparisons is the \textit{Allan deviation}, $\sigma_y(\tau)$, which characterises the fractional frequency instability of a clock or oscillator.
For example, an Allan deviation of $10^{-17}/\sqrt{\tau}$ means that after averaging for $\tau = 10{,}000$~s the clock's fractional frequency uncertainty is $10^{-19}$, i.e.\ it gains or loses less than one second in $10^{19}$~seconds (${\sim}300$~billion years).
When we state, for example, that a system achieves ``45~fs TDEV at 40{,}960~s,'' this means that when timing measurements are averaged over approximately 11 hours, the residual timing fluctuations are only 45 femtoseconds ($45 \times 10^{-15}$~s).

An important distinction that recurs throughout this paper is between \textit{frequency comparison} and \textit{time synchronisation}.
Frequency comparison determines whether two clocks tick at the same rate; it can be achieved by transmitting a continuous optical carrier and tracking its phase, without ever establishing a common time epoch.
Time synchronisation, by contrast, requires determining the absolute offset between two clocks, effectively answering ``what time is it there?'' rather than ``does your clock run at the same rate as mine?''
The two tasks have very different technical requirements: coherent optical frequency transfer already operates at the $10^{-19}$ fractional level over continental distances~\cite{Droste2013,Schioppo2022}, while direct time synchronisation lags by orders of magnitude (Section~\ref{sec:sync-gap}).

Another important concept is \textit{chromatic dispersion}: in optical fibres, different wavelengths of light travel at slightly different speeds, causing a short pulse to broaden as it propagates.
For classical time transfer, this broadening directly degrades the timing precision and must be carefully compensated.
Several quantum protocols discussed in this paper exploit a phenomenon called \textit{nonlocal dispersion cancellation}, where the frequency correlations between entangled photon pairs automatically eliminate this broadening~\cite{Giovannetti2001prl}, a property with no classical analogue.

With these concepts in hand, we now turn to the quantum protocols themselves, beginning with the earliest proposals and tracing their development through to the most recent experimental demonstrations.


\section{Overview of Quantum Protocols for Time Synchronisation and Distribution}\label{sec:protocols}

Over the past two decades, several families of quantum protocols have been proposed and, in some cases, experimentally demonstrated for clock synchronisation and time distribution.
These protocols pursue two different goals: higher timing precision through quantum correlations, and security guarantees that detect or prevent timing attacks that are invisible to classical systems.
The two value propositions are sometimes conflated, but they are independent: a protocol can offer one without the other.

In the following, we review the main protocol families, grouped by their physical principle.
For each family, we describe the underlying mechanism in accessible terms, summarise the key theoretical and experimental results, and assess the current maturity and suitability for practical deployment.
A comparative summary is provided in Section~\ref{sec:comparison}.

\subsection{Entanglement-Based Clock Synchronisation}\label{sec:entanglement-qcs}

The earliest quantum clock synchronisation (QCS) protocols aim to determine the time offset between two distant clocks using shared quantum states, without relying on knowledge of the signal travel time between the parties~\cite{Jozsa2000,Chuang2000}.
In a typical setup, a source distributes entangled qubit pairs so that each party holds one qubit.
Both parties independently perform local measurements and subsequently exchange only classical messages (measurement outcomes) over a standard communication channel.
From the correlations in their results, they compute the clock offset.
The key idea is that timing information is encoded in the quantum correlations of the shared entangled state, rather than in the arrival time of any signal.
This distinguishes these protocols from classical Einstein synchronisation, where precise knowledge of signal propagation delays is required.

The foundational proposal by Jozsa, Abrams, Dowling, and Williams~\cite{Jozsa2000} launched the field in 2000.
However, Preskill~\cite{Preskill2000} soon identified a subtle but serious issue: defining the shared entangled state implicitly requires a common phase reference between the two parties, which itself presupposes some degree of prior synchronisation.
This ``phase reference problem'' remained an open challenge for nearly two decades until Ilo-Okeke, Tessler, Dowling, and Byrnes~\cite{IloOkeke2018} demonstrated in 2018 that entanglement purification can eliminate the unknown phase errors, making these protocols viable in principle without pre-synchronised clocks.

From the theoretical perspective, Chuang~\cite{Chuang2000} showed that a quantum algorithm can determine $n$ bits of clock offset using only $O(n)$ quantum messages, compared to exponentially many classical messages, an exponential speedup in communication complexity.
However, no spatially separated experimental demonstration of foundational QCS has been achieved to date; the only validation is a nuclear magnetic resonance (NMR) simulation with co-located qubits~\cite{Zhang2004}.
These protocols are therefore foundational: they established the theoretical framework and demonstrated that quantum resources can, in principle, offer exponential advantages.
However, they are not yet candidates for practical deployment.

\subsection{Quantum-Enhanced Timing and Two-Way Time Transfer}\label{sec:qtwtt}

A second family of protocols takes a quantum metrology approach: instead of encoding timing in qubit correlations, it uses frequency-entangled photon pairs as ultra-precise timing signals~\cite{Giovannetti2001nature,Giovannetti2001prl}.
A source generates pairs of photons whose individual frequencies are uncertain, but whose sum is fixed precisely by energy conservation.
One photon is sent to each party, and the difference in recorded arrival times reveals the clock offset.

The quantum advantage is twofold.
First, with $N$ entangled photons, the timing precision scales as $1/N$ (the Heisenberg limit) rather than the classical $1/\sqrt{N}$ (shot-noise limit), meaning fewer photons are needed for the same precision.
Second, and perhaps more practically important, the frequency entanglement provides automatic dispersion cancellation: when one photon is broadened by chromatic dispersion in fibre or atmosphere, the correlation with its partner exactly compensates this broadening, to all orders~\cite{Giovannetti2001prl}.
The synchronisation precision is therefore insensitive to pulse distortion caused by the transmission medium, a significant advantage over classical time transfer, where dispersion is a major error source.
Valencia, Scarcelli, and Shih~\cite{Valencia2004} provided the first experimental demonstration in 2004, achieving picosecond-level resolution over 3~km of spooled optical fibre, although nonlocal dispersion cancellation was not implemented in this proof-of-principle experiment.

The practical evolution of this approach is quantum two-way time transfer (Q-TWTT), which combines the bidirectional geometry of classical two-way time transfer with nonlocal dispersion cancellation~\cite{Hou2019}.
In Q-TWTT, two spontaneous parametric down-conversion (SPDC) crystals, one at each end, generate frequency-entangled photon pairs.
The signal photon from each crystal traverses the long fibre link in opposite directions, while its partner (the idler) remains local.
At each end, the arriving signal photon and the local idler are detected, and the difference between their arrival times is recorded.
The clock offset is extracted from the combined two-way measurement, where symmetric propagation delays cancel (as in classical two-way time transfer).
The critical quantum advantage is that the frequency anti-correlation between signal and idler means the dispersive broadening of the signal photon is exactly compensated by the correlation measurement with its partner, nonlocally and to all orders of dispersion.

Q-TWTT has progressed rapidly from laboratory demonstrations to metropolitan-scale deployments.
Hou et al.~\cite{Hou2019} demonstrated Q-TWTT over 20~km of fibre, achieving a time deviation (TDEV) of 45~fs at 40{,}960~s averaging with an evaluated synchronisation uncertainty of 2.46~ps.
Hong et al.~\cite{Hong2022} extended the range to 50~km, reaching 54.6~fs TDEV at 57{,}300~s with a common reference clock and, importantly, 89.5~fs with independent clocks, confirming that the femtosecond stability is not an artefact of shared-clock correlations.
The same group~\cite{Hong2024} subsequently deployed Q-TWTT over 103~km of urban fibre with 38~dB attenuation, reaching 0.28~ps TDEV at 40{,}000~s with 13.9~ps evaluated uncertainty; the latter reflects residual systematic effects in a real-world deployment.
Most recently, Shi et al.~\cite{Shi2024} pushed the distance frontier to 250~km of direct fibre with approximately 55~dB total loss, maintaining 0.6~ps TDEV at 10{,}240~s.
Xiang et al.~\cite{Xiang2023} demonstrated Q-TWTT over a hybrid channel comprising 2~km of turbulent free-space and 7~km of deployed fibre, achieving 1.71~ps TDEV at 50~s and 144~fs at 6{,}400~s despite more than 25~dB loss, establishing the feasibility of Q-TWTT for space-ground links.
In a complementary effort, Lafler et al.~\cite{Lafler2024} demonstrated two-way quantum time transfer over a 1.6~km free-space testbed with software-emulated satellite motion and daytime background noise representative of a 700~km LEO link, reaching picosecond-scale precision with commercial off-the-shelf components.
The stability results are roughly three orders of magnitude better than coincidence-based methods, and the synchronisation uncertainties of a few to ${\sim}$14~ps represent the practical accuracy achievable today.
These results establish Q-TWTT as the best-performing entanglement-based approach for time distribution, with demonstrated suitability for metropolitan fibre, intercity fibre, and emerging free-space links.

\subsection{QKD-Based and Single-Photon Time Transfer}\label{sec:qkd-time}

A pragmatic family of protocols extracts timing information from the temporal correlations of photon pairs that are already used, or could be used, for QKD~\cite{Ho2009}.
An SPDC source produces time-correlated photon pairs; one photon travels to each party, and both record the arrival time stamped by their local clock.
By statistically cross-correlating the two detection-time series, the position of the coincidence peak yields the clock offset.
The precision is limited by detector timing jitter (typically 30--100~ps for commercial single-photon avalanche diodes) and the width of the biphoton correlation function, rather than by dispersion cancellation as in Q-TWTT.

The primary appeal of this approach is integration with quantum communication infrastructure.
Ho, Lamas-Linares, and Kurtsiefer~\cite{Ho2009} demonstrated in 2009 that QKD photon correlations can provide clock synchronisation without dedicated synchronisation hardware, relaxing the reference-clock accuracy requirement by approximately five orders of magnitude compared to previous schemes.
Lamas-Linares and Troupe~\cite{LamasLinares2018} later showed that this approach provides inherent security: to compromise timing, an adversary would need to perform high-probability quantum non-demolition measurements on single photons, an extremely challenging task with current or foreseeable technology.
Lee et al.~\cite{Lee2019} demonstrated bidirectional exchange of SPDC photon pairs with 51~ps precision, and their subsequent security analysis~\cite{Lee2019security} identified a practical vulnerability (asymmetric delay attacks) that any deployed system must address through additional countermeasures.

Two results stand out for practical deployment.
Dai et al.~\cite{Dai2020} demonstrated the feasibility of satellite-based quantum-secure time transfer using the Micius quantum satellite, achieving 30--60~ps precision (after averaging) over a quantum downlink channel combined with a classical uplink.
Tang et al.~\cite{Tang2023} demonstrated multi-user quantum clock synchronisation in an entanglement distribution network over 75~km of fibre, achieving 4.45~ps uncertainty and 426~fs TDEV at 4{,}000~s averaging using an SNSPD detector.
These results confirm that QKD-based time transfer is a natural add-on to quantum networks, providing timing as a byproduct of quantum communication with no additional hardware overhead.

It is worth noting that entanglement-based QKD architectures with an SPDC source at each end already contain the full hardware required for Q-TWTT (Section~\ref{sec:qtwtt}): each side possesses a local idler photon and receives a signal photon from the remote source, enabling dispersion-cancelled timing through the same photon stream used for key distribution.
Such dual-source QKD links could therefore provide both coincidence-based and dispersion-cancelled time transfer simultaneously, with the choice determined by post-processing rather than additional hardware.

\subsection{Hong-Ou-Mandel Interference-Based Synchronisation}\label{sec:hom-sync}

Hong-Ou-Mandel (HOM) interference provides what is arguably the sharpest timing reference available in quantum optics~\cite{Quan2016,Lyons2018}.
The principle relies on a distinctive quantum effect: when two indistinguishable photons arrive at a beam splitter simultaneously, quantum mechanics dictates that they always exit together through the same output port.
Coincidence detections (one photon in each output) therefore drop to zero when the photons arrive at the same time, producing a characteristic ``HOM dip'' as a function of relative delay.
The width of this dip is determined by the photon bandwidth and can be extremely narrow, providing a timing reference that reaches attosecond precision in principle.

For clock synchronisation, two sources of indistinguishable photons (one referenced to each clock) send photons to a central beam splitter.
The centre of the HOM dip reveals the clock offset.
The key advantage over coincidence-based methods (Section~\ref{sec:qkd-time}) is that HOM interference is sensitive to the quantum indistinguishability of the photons, not just their detection times, which provides intrinsically higher resolution.

Quan et al.~\cite{Quan2016} provided the first proof-of-concept demonstration over 4~km of spooled fibre, achieving 0.44~ps TDEV at 16{,}000~s averaging.
Lyons et al.~\cite{Lyons2018} then pushed HOM interferometry to its fundamental limits in a tabletop experiment, demonstrating 6~attosecond accuracy and 16~attosecond precision, two orders of magnitude beyond previous HOM approaches.
While the Lyons result is a laboratory measurement of path-length difference rather than a clock synchronisation over distance, it establishes the ultimate precision benchmark for what HOM-based timing could achieve.
The challenge for this approach remains scaling to longer distances, where photon loss and the requirement for indistinguishable photons from separate sources become increasingly demanding.

\subsection{Quantum Network of Clocks}\label{sec:network-clocks}

Rather than transferring a time signal between two clocks, the quantum network of clocks approach connects multiple atomic clocks across a network into a single collective quantum sensor~\cite{Komar2014}.
All clocks in the network are entangled into a shared quantum state, typically a Greenberger--Horne--Zeilinger (GHZ) state, which is a generalisation of a Bell pair to many parties.
Each clock makes its own local frequency measurement, and the results are combined.
Because the clocks share entanglement, their collective measurement achieves Heisenberg-limited precision: for $K$ clocks each with $N$ atoms, the frequency uncertainty scales as $1/(K \cdot N)$ rather than the classical $1/\sqrt{K \cdot N}$.

K{\'o}m{\'a}r et al.~\cite{Komar2014} proposed this ``world clock'' concept in 2014, showing that a network of 10 optical lattice clocks with 1{,}000 atoms each could in principle achieve an Allan deviation of $\sigma_y(\tau) \sim 2 \times 10^{-18}$ at 1~s, a stability exceeding what any single clock could provide.
The entanglement is distributed via photonic links: each clock node contains trapped atoms or ions that are entangled with a photon, and these photons are interfered at intermediate stations to create long-distance entanglement between nodes.

Nichol et al.~\cite{Nichol2022} provided the first experimental realisation in 2022, entangling two $^{88}$Sr$^{+}$ ion clocks separated by approximately 2~m via a photonic link.
For frequency comparisons between the two ions, entanglement reduced the required number of measurements by nearly the factor predicted for the Heisenberg limit.
More recently, Cao et al.~\cite{Cao2024} demonstrated GHZ states of up to 9 strontium clock qubits at a single site, a prerequisite for scaling to larger networks.

This approach represents the most ambitious long-term vision for quantum timekeeping: a distributed quantum clock whose stability surpasses any individual atomic clock.
However, scaling beyond proof-of-principle demonstrations requires quantum repeaters for long-distance entanglement distribution and remains a goal for the coming decade~\cite{Azuma2023}.

\subsection{Comparison and Assessment}\label{sec:comparison}

Table~\ref{tab:protocol-comparison} summarises the key experimental results across protocol families.
An important benchmark is the quantum-limited (but non-entangled) optical time transfer by Caldwell et al.~\cite{Caldwell2023}: synchronisation to 320~attoseconds over a 300~km folded free-space path using optical frequency comb pulses detected at the single-photon level, with TDEV scaling as 1.6~fs$/\sqrt{\tau}$.
This result uses a folded path geometry and a research-grade frequency comb that is not commercially available; it is a laboratory demonstration, not a deployable technology.
It does, however, establish the physical limits of what classical detection at the single-photon level can achieve.

\begin{table}[ht]
\centering
\caption{Comparison of experimental quantum time synchronisation results. The classical benchmark (Caldwell et al.) is included for reference.}
\label{tab:protocol-comparison}
\small
\begin{tabularx}{\textwidth}{l l l l l X}
\hline
\textbf{Protocol family} & \textbf{Stability (TDEV)} & \textbf{Uncertainty} & \textbf{Distance} & \textbf{Medium} & \textbf{Reference} \\
\hline
Entangled-photon (GLM) & ps-level & --- & 3 km & Fibre & Valencia et al.~\cite{Valencia2004} \\
Q-TWTT & 45 fs @ 40{,}960 s & 2.46 ps & 20 km & Fibre (spool) & Hou et al.~\cite{Hou2019} \\
Q-TWTT & 54.6 fs @ 57{,}300 s & 1.3 ps$^\dagger$ & 50 km & Fibre (spool) & Hong et al.~\cite{Hong2022} \\
Q-TWTT & 0.28 ps @ 40{,}000 s & 13.9 ps & 103 km & Urban fibre & Hong et al.~\cite{Hong2024} \\
Q-TWTT & 0.6 ps @ 10{,}240 s & --- & 250 km & Fibre & Shi et al.~\cite{Shi2024} \\
Q-TWTT (hybrid) & 144 fs @ 6{,}400 s & --- & 2+7 km & Free-space+fibre & Xiang et al.~\cite{Xiang2023} \\
Q-TWTT (sat.\ emul.) & ps-level & --- & 1.6 km & Free-space & Lafler et al.~\cite{Lafler2024} \\
QKD-based & --- & 30--60 ps & Satellite & Free-space & Dai et al.~\cite{Dai2020} \\
QKD-based (network) & 426 fs @ 4{,}000 s & 4.45 ps & 75 km & Fibre & Tang et al.~\cite{Tang2023} \\
HOM interference & --- & 6 as & Lab & Free-space & Lyons et al.~\cite{Lyons2018} \\
HOM interference & 0.44 ps @ 16{,}000 s & --- & 4 km & Fibre & Quan et al.~\cite{Quan2016} \\
Entangled clock network & Heisenberg scaling & --- & 2 m & Lab & Nichol et al.~\cite{Nichol2022} \\
\hline
\rowcolor{gray!15}
Classical benchmark & 1.6 fs$/\sqrt{\tau}$ & 320 as sync. & 300 km & Free-space & Caldwell et al.~\cite{Caldwell2023} \\
\hline
\multicolumn{6}{l}{\parbox{\textwidth}{\footnotesize $^\dagger$Mean time bias for high-spectral-consistency sources; not a full systematic uncertainty budget\\ (cf.\ Hou et al.\ 2.46~ps and Hong et al.\ 2024: 13.9~ps).}} \\
\end{tabularx}
\end{table}

Several observations follow from this comparison.
Q-TWTT currently offers the most mature combination of precision and practical distance, with femtosecond-level stability demonstrated over metropolitan fibre links up to 250~km; its primary advantage, nonlocal dispersion cancellation, addresses a real limitation of classical fibre-based time transfer.
QKD-based time transfer offers a different value proposition: rather than competing on raw precision, it provides timing as a natural byproduct of quantum communication infrastructure, with security benefits against certain classes of timing attacks.
HOM interference achieves the highest raw precision (attosecond level) but remains limited to short distances.
The case for quantum time transfer, therefore, rests on a portfolio of capabilities that no classical method can offer: dispersion-immune precision (Q-TWTT), integration with quantum communication networks (QKD-based), security against timing attacks, and, in the long term, Heisenberg-limited collective timekeeping through entangled clock networks.
No single quantum protocol delivers all four; rather, different protocol families address different needs, and can in some architectures be combined (Section~\ref{sec:qkd-time}).


\section{Practical Challenges and Opportunities}\label{sec:challenges}

The previous section presented the main families of quantum time synchronisation protocols and their experimental achievements.
However, a significant gap remains between the theoretical promise of these protocols and their practical performance.
This section examines the key bottlenecks that limit the performance of current quantum time synchronisation systems.

\subsection{From Theoretical Limits to Experimental Reality}\label{sec:theory-practice}

Two of the most frequently cited quantum advantages in time synchronisation are Heisenberg-limited precision scaling and nonlocal dispersion cancellation.
Both are real physical effects, but in practice they are substantially weaker than theoretical predictions suggest.

\textbf{Heisenberg scaling.}
The Heisenberg limit predicts that $N$ entangled particles can achieve a measurement precision scaling as $1/N$, compared to the classical $1/\sqrt{N}$ (Section~\ref{sec:network-clocks}).
In practice, Demkowicz-Dobrza{\'n}ski et al.~\cite{DemkowiczDobrzanski2012} proved that even small amounts of noise (photon loss, dephasing) reduce this scaling back to the standard quantum limit asymptotically.
The realistic quantum advantage becomes a constant factor improvement rather than a polynomial speedup.
The best experimental result to date, a spin-squeezed strontium clock at JILA~\cite{Yang2025}, achieves only approximately 2~dB beyond the standard quantum limit using ${\sim}30{,}000$ atoms, compared to the ${\sim}45$~dB that Heisenberg scaling would predict.
Entangled clock networks are not without value, but the realistic advantage is a modest constant-factor improvement rather than the dramatic scaling suggested by theoretical proposals.

\textbf{Dispersion cancellation.}
Nonlocal dispersion cancellation, the mechanism that gives Q-TWTT its primary advantage over classical fibre-based time transfer (Section~\ref{sec:qtwtt}), is a quantum effect with no classical analogue~\cite{Giovannetti2001prl}.
Experimental results confirm the effect but show that the cancellation is partial rather than complete: 
measurements over deployed fibre show a roughly 2--3-fold reduction in coincidence peak broadening~\cite{Chua2022}, while Yu et al.~\cite{Yu2025} observed a fivefold improvement in QKD key rate attributable to dispersion resilience.
The discrepancy arises from residual effects that energy-time entanglement does not address, including polarisation mode dispersion, finite pump bandwidth, and detector timing jitter~\cite{Hou2019,Hong2024}.
Dispersion cancellation is therefore a real but bounded advantage, most relevant for long fibre links (above ${\sim}50$~km) where chromatic dispersion dominates the timing error budget.

\subsection{Hardware Bottlenecks}\label{sec:hardware}

Three hardware components dominate the performance limitations of current quantum time synchronisation systems: the photon source, the single-photon detectors, and the supporting infrastructure (cryogenics, time-tagging electronics).

\textbf{Photon sources.}
Nearly all QTS experiments to date use SPDC sources, typically pumped by a continuous-wave laser, which generate photon pairs probabilistically at rates ranging from ${\sim}1$ to ${\sim}100$~Mpairs/s depending on the pump power and crystal design.
This probabilistic nature creates a fundamental trade-off: increasing the pump power raises the pair-generation rate but also increases the probability of generating multiple pairs simultaneously.
These multi-pair events introduce accidental coincidences that bias the timing measurement and degrade precision.
In practice, SPDC sources are operated at low mean pair numbers ($\mu \approx 0.01$--$0.05$ per pulse) to keep multi-pair contamination below a few per cent, which limits the usable coincidence rate to tens or hundreds of kilohertz.
Deterministic single-photon sources based on quantum dots could, in principle, overcome this trade-off, but they currently require cryogenic cooling (${\sim}2$--5~K), produce single photons rather than entangled pairs (requiring additional entanglement generation steps), and are not yet competitive at telecom wavelengths.
Integrated photonic sources with engineered spectral properties represent a more near-term path to improved performance.

\textbf{Single-photon detectors.}
For correlation-based QTS protocols, including Q-TWTT and QKD-based time transfer (Sections~\ref{sec:qtwtt} and~\ref{sec:qkd-time}), detector timing jitter is the single largest bottleneck for precision.
HOM-based synchronisation (Section~\ref{sec:hom-sync}) is a notable exception: because the timing information is encoded in the two-photon interference dip rather than in individual detection timestamps, the precision is determined by the photon bandwidth and is largely insensitive to detector jitter, although this advantage has not yet been demonstrated over practical distances.
The combined system jitter from two detectors directly sets the minimum width of the coincidence peak from which the clock offset is extracted.
Superconducting nanowire single-photon detectors (SNSPDs) achieve the best jitter performance, with values as low as ${\sim}3$--20~ps at telecom wavelengths, along with detection efficiencies exceeding 95\%.
However, SNSPDs require cryogenic cooling to ${\sim}0.8$--2.5~K, which adds approximately 1--1.5~kW of electrical power, 30--100~kg of weight, and 100{,}000~EUR of cost per multi-channel system.
The alternative, InGaAs single-photon avalanche diodes (SPADs), operate near room temperature at a fraction of the cost but with 5--10$\times$ worse jitter (${\sim}40$--170~ps) and lower efficiency (30--40\% in gated mode with a pulsed source, below 20\% in free-running mode).
Moreover, SPAD timing jitter degrades as the detection rate approaches saturation, so simply increasing the source brightness to accumulate coincidences faster is counterproductive: the resulting jitter increase offsets the statistical gain, imposing a practical ceiling on achievable QTS precision with these detectors.
The practical consequence is significant\footnote{The combined jitter of two identical detectors is $\sigma_t = \sigma_\mathrm{det}\sqrt{2}$; the mean offset uncertainty after $N$ coincidences is $\sigma_\mathrm{mean} = \sigma_t / \sqrt{N}$, so the required number of coincidences scales as $N = \sigma_t^2 / \sigma_\mathrm{target}^2$.}: reaching 1~ps timing precision with SNSPDs at 20~ps jitter requires approximately 800 coincidences, while achieving the same precision with InGaAs SPADs at 76~ps jitter requires approximately 11{,}500 coincidences, roughly 14$\times$ more measurement time from jitter alone.

An emerging alternative, MgB$_2$ SNSPDs operating at 20~K~\cite{Charaev2024}, could reduce the cryogenic burden by enabling simpler and cheaper cooling systems, although they have not yet been demonstrated to match the performance of conventional SNSPDs.

\textbf{System complexity.}
Beyond the core photon source and detectors, a complete QTS system requires precision time-tagging electronics (with sub-picosecond resolution to avoid degrading the detector performance), polarisation stabilisation, and, in many cases, dispersion compensation modules.
The total size, weight, power, and cost (SWaP-C) of a quantum time synchronisation node is estimated at 2--3$\times$ that of a comparable classical optical time transfer system, with the cryogenic detectors accounting for the largest share of the overhead.

\subsection{Distance and Channel Limitations}\label{sec:distance}

The maximum distance over which quantum time synchronisation can operate is ultimately set by photon loss in the transmission channel.

\textbf{Fibre loss.}
Standard telecom fibre attenuates at approximately 0.2~dB/km at 1550~nm, meaning that a photon has only a 10\% chance of surviving a 50~km link, 1\% for 100~km, and $10^{-5}$ for 250~km.
For protocols that require coincidence detection of photon pairs, the loss is doubly punishing: pair survival probability scales as the product of both channel transmissions.
In practice, deployed fibre suffers additional loss from splices, connectors, and bending, so that the 103~km urban Q-TWTT experiment by Hong et al.~\cite{Hong2024} experienced 38~dB total attenuation, equivalent to approximately 190~km of ideal fibre.
Neumann et al.~\cite{Neumann2022} demonstrated entanglement distribution over 248~km of deployed transnational fibre but achieved only 9 detected pairs per second, a rate that, while sufficient for entanglement verification, is too low for high-precision time synchronisation.

Without quantum repeaters, the practical distance limit for entanglement-based time synchronisation over deployed fibre is approximately 200--300~km.
Even recent advances in hollow-core fibre, which reduce attenuation to approximately 0.09~dB/km, do not fundamentally change this picture because the loss remains exponential.

\textbf{Environmental effects.}
Deployed fibre is subject to temperature-induced delay fluctuations of approximately 40~ps/(km$\cdot$K).
Over a 100~km link, a 1~K temperature change causes a 4~ns delay shift, four orders of magnitude larger than the sub-picosecond precision targets of advanced QTS systems.
Two-way protocols such as Q-TWTT cancel this effect to first order (since the common-mode delay affects both directions equally), which is one of their primary practical advantages.
The transition from laboratory to field consistently degrades QTS performance: a recent metropolitan deployment in Stockholm~\cite{Alqedra2025} achieved approximately 100~ps precision over 20~km of deployed fibre, roughly 100$\times$ worse than comparable laboratory results over similar distances.

\textbf{Satellite channels.}
Free-space satellite links enable longer distances but at lower precision.
The Micius demonstration by Dai et al.~\cite{Dai2020} achieved 30--60~ps timing precision over a quantum satellite downlink, roughly 100$\times$ worse than the best fibre-based Q-TWTT results, limited by atmospheric turbulence, lower count rates, and brief satellite pass durations (typically a few minutes for low-Earth-orbit satellites).
Satellite-based QTS also depends on weather conditions, as cloud cover blocks single-photon signals.

\textbf{Quantum repeaters and range extension.}
To extend entanglement-based QTS beyond ${\sim}300$~km, quantum repeaters, devices that store and relay quantum states at intermediate nodes, are the standard prescription for overcoming exponential fibre loss.
However, quantum repeaters are not straightforwardly compatible with QTS protocols, particularly Q-TWTT.

The core problem is a bandwidth mismatch.
Q-TWTT achieves its dispersion-cancellation advantage because the entangled photon pairs exhibit very broad frequency correlations (typically ${\sim}1$~THz bandwidth), which enable tight picosecond-level timing correlations.
Quantum memories, which must temporarily store quantum states at each repeater node, accept only a narrow slice of this bandwidth (${\sim}1$--10~GHz at best), effectively filtering out the very correlations that make Q-TWTT useful~\cite{Vitullo2018}.
This would broaden the timing correlation by a factor of 100--1{,}000, defeating the purpose of sub-picosecond synchronisation.
Additionally, quantum memories introduce variable storage delays with readout timing jitter of 50--500~ps, far exceeding QTS precision targets.

Rather than waiting for quantum repeaters, the leading Q-TWTT research group has pursued a practical alternative: Hong et al.~\cite{Hong2026cascaded} demonstrated cascaded quantum time transfer over $2 \times 100$~km using relay stations that generate \textit{fresh} entangled photon pairs at each hop, achieving 3.82~ps TDEV at 10~s and 0.39~ps at 5{,}120~s.
This approach preserves the broadband correlations on each segment but requires each relay node to be physically secured (a trusted-node architecture).

No multi-segment quantum repeater chain has been demonstrated to date for any application, and operational repeater networks are estimated at the mid-2030s or later~\cite{Azuma2023}.
The additional constraints imposed by QTS make repeater-assisted quantum timing even more distant.

\subsection{Security of Timing: Focus on QKD-Based Protocols}\label{sec:security}

Security against timing manipulation is the most distinctive advantage that quantum time protocols offer over classical methods.
Classical time transfer, whether based on GPS, Network Time Protocol (NTP), or precision optical links, is inherently vulnerable to spoofing and delay attacks.
An adversary who can intercept and retransmit a classical timing signal can introduce an undetectable timing offset.
In domains where timing integrity is critical, e.g.\ financial trading, power grid synchronisation, or military navigation, such attacks represent a serious and growing threat.

QKD-based time transfer protocols (Section~\ref{sec:qkd-time}) address this vulnerability by encoding timing information in the quantum correlations of single photons.
Lamas-Linares and Troupe~\cite{LamasLinares2018} showed that to compromise the timing without detection, an adversary would need to perform quantum non-demolition measurements of individual photons with high probability, a task that is extremely challenging with current or foreseeable technology.
The quantum origin of the timing correlations can be verified through standard entanglement witnesses or Bell-state measurements: the Stockholm metropolitan demonstration~\cite{Alqedra2025}, for instance, confirmed a fidelity of 0.817 to a maximally entangled Bell state, providing cryptographic assurance that the timing signal has not been tampered with.

However, the security of QKD-based time transfer is not unconditional.
Lee et al.~\cite{Lee2019security} identified a practical vulnerability: an adversary can introduce different delays in the two directions of a bidirectional protocol (an asymmetric delay attack) without being detected by standard coincidence measurements.
Multi-photon emissions from practical SPDC sources create additional attack surfaces, although timing information (carried by photon arrival times rather than polarisation states) is inherently less vulnerable to photon-number-splitting attacks than QKD key bits.
Side-channel attacks developed against QKD hardware (such as detector blinding with bright light) are directly applicable to QTS systems that share the same components.

For example, Miller~\cite{Miller2025} showed that timing desynchronisation among the Micius satellite's laser diodes allowed signal and decoy states to be distinguished with ${\sim}$98.7\% probability, breaking the decoy-state security guarantee.
This class of vulnerability applies equally to QTS: if photon states carry distinguishable signatures in non-operational degrees of freedom, all security advantages of quantum timing are lost.

In this context, the security advantage of quantum time transfer is real and significant, but it requires careful implementation.
Countermeasures such as authenticated classical channels, delay monitoring, and device-independent verification protocols add complexity but are essential for a deployable system.
The key insight is that QKD-based protocols provide security as an inherent physical property of the quantum channel, not as a computational assumption that could be broken by future algorithms.
This distinction is particularly relevant given the advancing threat of quantum computers to classical cryptographic infrastructure.


\section{Classical Time Synchronisation: The Existing Landscape}\label{sec:classical-protocols}

Quantum time synchronisation protocols must be evaluated against a mature and rapidly improving ecosystem of classical methods.
This section surveys the principal classical techniques to establish the baseline against which quantum approaches must demonstrate added value.

\subsection{Network Protocols: NTP, PTP, and White Rabbit}\label{sec:network-sync}

The vast majority of clock synchronisation worldwide relies on network protocols.
The NTP, introduced in 1985~\cite{Mills1991}, synchronises clocks over the public Internet with a typical accuracy of 1--10~ms over wide-area networks, improving to approximately 1~$\mu$s on local networks with GPS pulse-per-second references.
NTP is deployed on virtually every networked device; modern versions support authentication via NTS (based on TLS), which protects against spoofing, but remain inherently vulnerable to delay manipulation attacks.

The Precision Time Protocol (PTP, IEEE 1588), first standardised in 2002, improves on NTP by using hardware timestamping at the physical layer.
With dedicated hardware, PTP achieves sub-10~ns synchronisation over local area networks and is the standard for telecommunications (5G synchronisation), financial trading, and industrial automation.

White Rabbit~\cite{Moreira2009,Lipinski2011}, developed at CERN and standardised as the IEEE 1588-2019 High Accuracy profile, extends PTP with Synchronous Ethernet for physical-layer frequency locking and a digital phase detector for fine phase measurement.
Lipi{\'n}ski et al.~\cite{Lipinski2011} demonstrated approximately 200~ps accuracy and 6~ps precision over a 5~km fibre link.
Dierikx et al.~\cite{Dierikx2016} demonstrated $\pm$2~ns stability over a 950~km fibre link, and Amies-King and Lucamarini~\cite{AmiesKing2025} recently achieved 4.0~ps TDEV at 400~s over 300~km of unrepeated fibre.
White Rabbit is fully open-source, commercially available from multiple vendors, and deployed at facilities ranging from CERN to national metrology institutes.
It represents the current practical standard for high-precision fibre-based time distribution.

It is worth noting, however, that White Rabbit inherits two distinct security limitations from PTP.
First, its authentication relies on MACsec or similar cryptographic primitives based on computational assumptions; this can, in principle, be addressed by upgrading to post-quantum cryptography or universal-hash-function-based message authentication.
The more fundamental vulnerability is that an adversary with physical access to the fibre link can manipulate the propagation delay itself, intercepting, delaying, and retransmitting signals without violating any authentication mechanism, since the attack targets the physical layer rather than the data content.
No cryptographic upgrade can close this gap, because the protocol has no means to verify that the propagation delay has not been tampered with.
For most current deployments, this is acceptable, but for adversarial environments where physical-layer timing integrity is required, it represents a vulnerability that quantum approaches could address.

\subsection{GNSS-Based Time Transfer}\label{sec:gnss}

Global navigation satellite systems (GNSS), particularly GPS, provide time transfer as a byproduct of their positioning function.
A GPS timing receiver at a known location can extract time referenced to UTC with an accuracy of approximately 2--10~ns.
GPS-disciplined oscillators (GPSDOs), which steer a local oscillator to the GPS signal, are the most common high-accuracy time references in commercial and infrastructure applications.

For higher-precision comparisons between remote clocks, the common-view technique has two stations observe the same satellite simultaneously, thereby cancelling satellite clock errors and most atmospheric delays, yielding an uncertainty of approximately 1--2~ns.
This is the technique used by the Bureau International des Poids et Mesures (BIPM) for computing Coordinated Universal Time (UTC).
Carrier-phase methods exploit the shorter wavelength of the GPS carrier signal (${\sim}19$~cm at L1) rather than the pseudorandom noise code, and the Differential Precise Time (DPT) technique has demonstrated approximately 20~ps intra-day accuracy with frequency stability reaching the sub-$10^{-17}$ level at week-long averaging.

The key advantage of GNSS-based methods is global coverage with inexpensive receivers.
Their key vulnerability is susceptibility to jamming and spoofing: GPS signals are weak (${\sim}$--160~dBW at the receiver), and commercially available jammers can deny or falsify GPS timing over wide areas.

\subsection{Two-Way Satellite Time and Frequency Transfer}\label{sec:twstft}

Two-Way Satellite Time and Frequency Transfer (TWSTFT) is the highest-accuracy satellite-based timing technique in operational use~\cite{Kirchner1991}.
Two ground stations simultaneously exchange coded timing signals via a geostationary satellite; by differencing the two one-way measurements, the satellite delay and most path-related errors cancel due to the geometry's symmetry.
This is the same two-way principle exploited by Q-TWTT (Section~\ref{sec:qtwtt}) in the quantum domain.
Modern TWSTFT using software-defined radio modems achieves an uncertainty of approximately 0.3~ns for intercontinental time transfer.
Together with GNSS, TWSTFT is one of the two primary techniques used by the BIPM to compute UTC/TAI.
Like other classical approaches, TWSTFT is vulnerable to signal spoofing and delay attacks, although the two-way geometry makes symmetric delay attacks harder to execute undetected.

\subsection{Fibre-Based Precision Time and Frequency Transfer}\label{sec:fibre-classical}

The highest-precision classical time and frequency transfer is achieved over optical fibre, using either actively stabilised time links or coherent optical frequency transfer.

\textbf{Actively stabilised time links.}
The ELSTAB system, developed at AGH University of Science and Technology in Poland, uses a two-way measurement scheme combined with electronically controlled delay lines to actively compensate temperature-induced propagation delay fluctuations~\cite{Krehlik2012,Sliwczynski2013}.
Krehlik et al.~\cite{Krehlik2012} achieved 0.3~ps TDEV at $10^5$~s averaging over a 60~km urban fibre loop, and {\'S}liwczy{\'n}ski et al.~\cite{Sliwczynski2013} demonstrated sub-picosecond time transfer over 420~km of deployed fibre.
These systems require no single-photon detectors or cryogenic equipment, making them significantly simpler to deploy than quantum alternatives with comparable precision.

\textbf{Coherent optical frequency transfer.}
For frequency comparison between the most precise atomic clocks, coherent optical transfer sends an ultrastable laser signal through telecom fibre and actively compensates fibre-induced phase noise.
Droste et al.~\cite{Droste2013} demonstrated frequency transfer at the $4 \times 10^{-19}$ level at 100~s over a single-span 1840~km fibre link, and operational networks now exist at continental scale: the REFIMEVE+ network in France spans over 2600~km~\cite{Lisdat2016}, and Schioppo et al.~\cite{Schioppo2022} compared ultrastable lasers at $7 \times 10^{-17}$ instability through a 2220~km network connecting the UK, France, and Germany.
It is worth noting that coherent optical transfer provides \textit{frequency} comparison (whether two clocks tick at the same rate) rather than direct \textit{time} transfer (establishing a common time epoch via a pulse-per-second signal).

\textbf{Quantum-limited optical time transfer.}
Caldwell et al.~\cite{Caldwell2023} demonstrated optical time transfer achieving 320~attosecond synchronisation over a 300~km free-space path, with TDEV scaling as 1.6~fs$/\sqrt{\tau}$, using optical frequency comb pulses detected at the single-photon level.
Several caveats are important for a fair comparison with quantum methods.
The 300~km distance was achieved using a folded-path geometry (transmitter and receiver co-located at the Mauna Loa Observatory, Hawaii, with a retroreflector on Haleakal\=a at ${\sim}$150~km), rather than a point-to-point deployed link.
The system requires a time-programmable optical frequency comb, a research instrument developed at NIST that is not commercially available.
This result is a laboratory-grade demonstration comparable in maturity to many of the quantum results discussed in Section~\ref{sec:protocols}.

\subsection{Summary}\label{sec:classical-summary}

\begin{table}[htb!]
\centering
\caption{Summary of classical time synchronisation protocols. Stability and uncertainty figures represent the best demonstrated results. Deployment status reflects operational (non-research) use. Coherent optical transfer provides frequency comparison only (not direct time transfer).}
\label{tab:classical-comparison}
\small
\begin{tabularx}{\textwidth}{l l l l l X}
\hline
\textbf{Protocol} & \textbf{Stability (TDEV)} & \textbf{Uncertainty} & \textbf{Distance} & \textbf{Medium} & \textbf{Reference / Status} \\
\hline
NTP & --- & ${\sim}1$~$\mu$s & Global & Internet & Mills~\cite{Mills1991}; ubiquitous \\
PTP (IEEE 1588) & --- & $<$10~ns & LAN & Ethernet & IEEE 1588-2019; widely deployed \\
White Rabbit & 4~ps @ 400~s & ${\sim}200$~ps & 300~km & Fibre & Amies-King~\cite{AmiesKing2025}; deployed \\
White Rabbit & --- & $\pm$2~ns & 950~km & Fibre & Dierikx~\cite{Dierikx2016}; deployed \\
GPS timing & --- & ${\sim}2$~ns & Global & Satellite & GPS SPS Standard; ubiquitous \\
Common-view GPS & --- & ${\sim}1$--2~ns & Global & Satellite & BIPM; operational \\
Carrier-phase GPS & --- & ${\sim}20$~ps & Global & Satellite & DPT technique; operational \\
TWSTFT & --- & ${\sim}0.3$~ns & Intercont. & Satellite & Kirchner~\cite{Kirchner1991}; operational \\
ELSTAB & 0.3~ps @ $10^5$~s & $<$50~ps & 60~km & Fibre & Krehlik~\cite{Krehlik2012}; deployed \\
ELSTAB & sub-ps & --- & 420~km & Fibre & {\'S}liwczy{\'n}ski~\cite{Sliwczynski2013}; deployed \\
Coherent optical & $\sigma_y$: $4 \times 10^{-19}$ & --- & 1840~km & Fibre & Droste~\cite{Droste2013}; research \\
\hline
\rowcolor{gray!15}
Caldwell et al. & 1.6~fs$/\sqrt{\tau}$ & 320~as sync. & 300~km$^*$ & Free-space & \cite{Caldwell2023}; research demo \\
\hline
\multicolumn{6}{l}{\footnotesize $^*$Folded path geometry (co-located transmitter/receiver with distant retroreflector), not a deployed link.} \\
\end{tabularx}
\end{table}

Table~\ref{tab:classical-comparison} summarises the classical landscape using metrics comparable to the quantum results in Table~\ref{tab:protocol-comparison}.
The most relevant benchmarks for quantum protocols depend on the application: for network synchronisation, White Rabbit (${\sim}200$~ps accuracy, commercially deployed) sets the practical standard; for satellite timing, GNSS and TWSTFT (${\sim}0.3$--2~ns) are operationally dominant; for metrology-grade fibre links, ELSTAB (${\sim}$ps TDEV over hundreds of kilometres) is the direct competitor.


\section{Quantum vs Classical: A Comparative Assessment}\label{sec:comparison-qvc}

Having described both quantum and classical approaches, this section provides a direct comparison, focusing on the practical question: under what conditions do quantum protocols offer an advantage that justifies their additional complexity and cost?

\subsection{Comparing Like with Like}\label{sec:fair-comparison}

A fair comparison must account for two factors that are often overlooked.

First, the maturity gap between quantum and classical systems is large.
Classical fibre-based time transfer (White Rabbit, ELSTAB) has been refined over more than a decade of field deployments, with well-characterised systematic errors and commercially available hardware.
Quantum time synchronisation, by contrast, is largely at the laboratory demonstration stage, with only a handful of field deployments; the Stockholm metropolitan demonstration~\cite{Alqedra2025} is among the first.
Performance figures from laboratory experiments do not directly predict field performance; as noted in Section~\ref{sec:distance}, the lab-to-field degradation for quantum systems has been approximately 100$\times$ in the published demonstrations.

Second, the most frequently cited classical benchmark, Caldwell et al.~\cite{Caldwell2023} at 320~as over 300~km, is itself a laboratory-grade demonstration using non-commercial research equipment over a folded free-space path (Section~\ref{sec:fibre-classical}).
Comparing it with quantum demonstrations at face value would be misleading, because neither system is deployable today.
The operationally relevant classical benchmarks are White Rabbit (${\sim}$200~ps accuracy, commercially deployed) for fibre and GNSS/TWSTFT (${\sim}$0.3--2~ns) for satellite-based applications.

\subsection{Where Classical Methods Are Sufficient}\label{sec:classical-sufficient}

For many applications, classical methods already provide more than adequate timing:

\begin{itemize}[nosep]
    \item \textbf{Telecommunications and data centres:} 5G networks require synchronisation to ${\sim}$1.5~$\mu$s (phase) and ${\sim}$30~ns (time), well within the capability of PTP and GPS. Even the strictest telecom profiles are satisfied by White Rabbit.
    \item \textbf{Financial trading:} MiFID~II regulations in the EU require timestamps accurate to 100~$\mu$s for most instruments and 1~$\mu$s for high-frequency trading. PTP with GPS backup meets these requirements.
    \item \textbf{Power grid synchronisation:} Synchrophasor measurements require ${\sim}$1~$\mu$s accuracy, achievable with GPS or PTP.
    \item \textbf{Scientific facilities:} Large-scale experiments (particle physics, radio astronomy) routinely use White Rabbit for sub-nanosecond synchronisation across kilometres of fibre.
\end{itemize}

In these scenarios, the sub-picosecond precision offered by quantum protocols is unnecessary, and the additional hardware complexity (SPDC sources, single-photon detectors, cryogenics) would add cost without practical benefit.

\subsection{Where Quantum Protocols Add Value}\label{sec:quantum-value}

Quantum time synchronisation adds value in specific scenarios that classical methods cannot address:

\textbf{Security against timing attacks.}
Classical timing protocols (NTP, PTP, GPS, and even TWSTFT) are vulnerable to spoofing and delay manipulation.
GPS jamming and spoofing devices are widely available and have been used to disrupt navigation and timing in both civilian and military contexts.
NTP and PTP lack built-in authentication (NTS for NTP and MACsec for PTP are being adopted, but rely on computational assumptions).
QKD-based time transfer provides timing authentication rooted in quantum physics rather than computational hardness: the presence of entanglement in the timing channel can be continuously verified, and any attempt to intercept and retransmit the timing signal disturbs the quantum state in a detectable way (Section~\ref{sec:security}).
For applications where timing integrity must be guaranteed against a sophisticated adversary, e.g.\ military operations, critical infrastructure, or high-value financial systems, this physical-layer security is a capability that no classical protocol can replicate.

\textbf{Dispersion-immune fibre time transfer.}
Over long fibre links ($>$50~km), chromatic dispersion is a significant error source for classical time transfer.
Classical systems compensate dispersion using hardware modules that add loss and must be calibrated for each link, or by operating in the O-band (near the zero-dispersion wavelength) at the cost of higher fibre attenuation.
Q-TWTT's nonlocal dispersion cancellation avoids both penalties, providing a 2--5$\times$ improvement in timing broadening without additional hardware or operating-wavelength constraints (Section~\ref{sec:theory-practice}).
This advantage is most relevant for metropolitan and intercity fibre links where sub-picosecond precision is required, and dispersion compensation is a limiting factor, for instance, in comparisons between national metrology institutes or synchronisation of distributed scientific instruments.

\textbf{Integration with quantum networks.}
As quantum communication networks are deployed for QKD and other quantum information tasks, timing synchronisation can be derived from the same photon correlations without additional hardware~\cite{Spiess2023,Tang2023}.
This dual-use capability means that organisations investing in quantum networks for security reasons receive timing as a byproduct.
The synchronisation precision achievable (${\sim}$4--30~ps) exceeds the requirements of most telecommunications and infrastructure applications, and the security guarantees come at no marginal cost.
This is not a performance advantage over classical timing but rather an economic and architectural advantage for organisations that are deploying quantum infrastructure for other purposes.

\textbf{Future: distributed quantum sensing.}
In the longer term, entangled clock networks (Section~\ref{sec:network-clocks}) could enable collective timekeeping with precision exceeding any individual clock.
This capability has no classical analogue: classical clocks can be compared and averaged, but entanglement allows measurements that exploit quantum correlations between distant clocks.
While this remains a decade or more away from practical realisation, it represents a fundamentally new capability, not merely an improvement over classical methods.

Quantum and classical time synchronisation are complementary rather than competing.
Classical methods provide the performance, range, and maturity needed for the vast majority of current timing applications.
Quantum methods add capabilities that classical systems cannot provide: physical-layer security against timing attacks, dispersion immunity without hardware compensation, and, in the longer term, Heisenberg-limited collective timekeeping.
The near-term value proposition is strongest in two scenarios: (1)~as a security layer for timing-critical infrastructure that faces adversarial threats, and (2)~as a low-cost addition to quantum communication networks being deployed for other purposes.


\section{Optical Clocks and the Synchronisation Gap}\label{sec:clocks}

Optical atomic clocks have improved by more than three orders of magnitude over the past two decades, with several systems now reaching fractional frequency uncertainties at or below $10^{-18}$.
Two architectures dominate: optical lattice clocks, which interrogate thousands of neutral atoms trapped in an optical standing wave and offer rapid averaging (stability of ${\sim}10^{-17}/\sqrt{\tau}$), and single-ion clocks, which achieve the lowest systematic uncertainties by probing a single trapped atom at the cost of slower averaging (${\sim}10^{-15}/\sqrt{\tau}$).
As of early 2026, the NIST $^{27}$Al$^{+}$ quantum logic clock holds the record at $5.5 \times 10^{-19}$~\cite{Marshall2025}, followed by the VTT MIKES $^{88}$Sr$^{+}$ clock at $7.9 \times 10^{-19}$~\cite{Lindvall2025} and the JILA $^{87}$Sr lattice clock at $8.1 \times 10^{-19}$~\cite{Aeppli2024}.
Table~\ref{tab:clock-records} summarises the leading results; notably, ion and lattice architectures have converged to comparable systematic uncertainties, while multi-ion approaches such as the PTB $^{115}$In$^{+}$ Coulomb crystal clock~\cite{Hausser2025} are beginning to close the stability gap.
Outside the laboratory, transportable optical clocks have reached the low $10^{-18}$ level: the PTB transportable $^{87}$Sr lattice clock recently demonstrated $2.1 \times 10^{-18}$ uncertainty during a field campaign~\cite{Nosske2025}, roughly a factor of four above the best laboratory systems, with the degradation arising from less controlled thermal environments, vibrations, and limited averaging time.

\begin{table}[ht]
\centering
\caption{Leading optical clock systematic uncertainties as of early 2026. Transportable results are shown separately.}
\label{tab:clock-records}
\small
\begin{tabularx}{\textwidth}{l l l l l X}
\hline
\textbf{Species} & \textbf{Type} & \textbf{Systematic unc.} & \textbf{Lab} & \textbf{Year} & \textbf{Reference} \\
\hline
$^{27}$Al$^{+}$ & Ion & $5.5 \times 10^{-19}$ & NIST & 2025 & Marshall et al.~\cite{Marshall2025} \\
$^{88}$Sr$^{+}$ & Ion & $7.9 \times 10^{-19}$ & VTT MIKES & 2025 & Lindvall et al.~\cite{Lindvall2025} \\
$^{87}$Sr & Lattice & $8.1 \times 10^{-19}$ & JILA & 2024 & Aeppli et al.~\cite{Aeppli2024} \\
$^{171}$Yb & Lattice & $1.4 \times 10^{-18}$ & NIST & 2018 & McGrew et al.~\cite{McGrew2018} \\
$^{115}$In$^{+}$ & Ion (multi) & $2.5 \times 10^{-18}$ & PTB & 2025 & Hausser et al.~\cite{Hausser2025} \\
$^{171}$Yb$^{+}$ (E3) & Ion & $3.2 \times 10^{-18}$ & PTB & 2016 & Huntemann et al.~\cite{Huntemann2016} \\
\hline
\rowcolor{gray!15}
$^{87}$Sr (transp.) & Lattice & $2.1 \times 10^{-18}$ & PTB & 2025 & Nosske et al.~\cite{Nosske2025} \\
\rowcolor{gray!15}
$^{87}$Sr (transp.) & Lattice & $5.5 \times 10^{-18}$ & RIKEN & 2020 & Takamoto et al.~\cite{Takamoto2020} \\
\hline
\end{tabularx}
\end{table}

\subsection{Can We Synchronise These Clocks?}\label{sec:sync-gap}

The rapid improvement in clock precision has widened the gap between what clocks can measure and what time transfer links can deliver.
A clock with fractional frequency uncertainty $\delta f/f$ requires a synchronisation link whose timing uncertainty $\delta t$ satisfies $\delta t < (\delta f/f) \times \tau$, where $\tau$ is the averaging time.
For a $10^{-19}$ laboratory clock averaged over $\tau = 10{,}000$~s, this demands sub-femtosecond synchronisation, a target that no demonstrated time transfer system (as opposed to frequency transfer) has reached over practical distances.
For field-deployed clocks at ${\sim}10^{-17}$, the requirement relaxes to ${\sim}100$~fs (0.1~ps), still below the synchronisation uncertainty of any current quantum method (the best Q-TWTT result is 2.46~ps, roughly 25$\times$ above this target), though Q-TWTT stability (TDEV) has already reached 45~fs.

\begin{figure}[htb!]
\centering
\includegraphics[width=\textwidth]{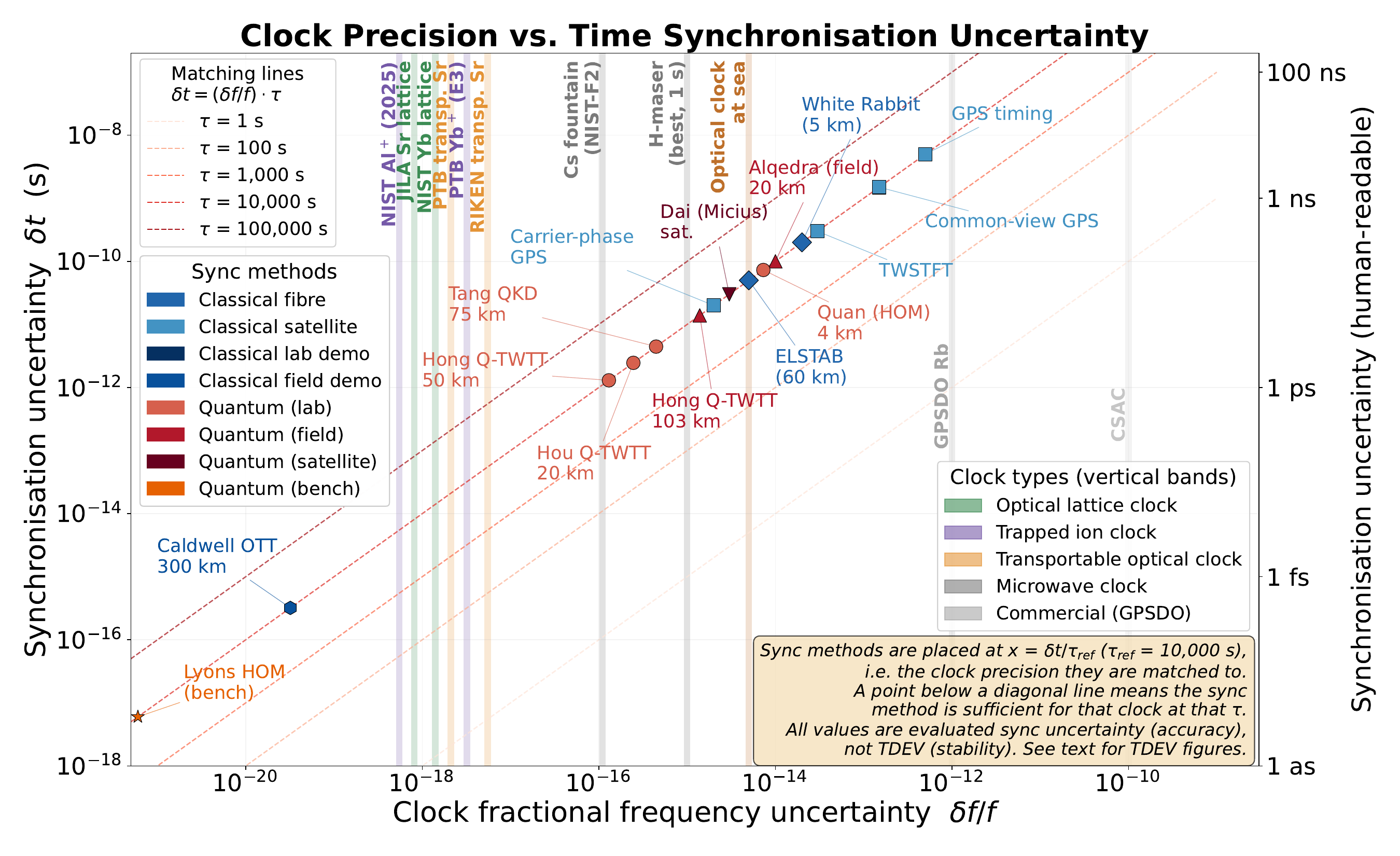}
\caption{Clock fractional frequency uncertainty versus synchronisation uncertainty for current time transfer methods.
Vertical bands indicate clock precision; scatter points show synchronisation methods placed at the clock precision they can serve at $\tau = 10{,}000$~s (i.e.\ $x = \delta t / \tau$).
Diagonal dashed lines mark the matching condition $\delta t = (\delta f/f) \cdot \tau$ for several averaging times.
A point below a line indicates the method is sufficient for that clock at that $\tau$.
Data are compiled from the experimental results reviewed in Sections~\ref{sec:protocols}--\ref{sec:classical-protocols} and the clock records in Table~\ref{tab:clock-records}.}
\label{fig:clock-vs-sync}
\end{figure}

Figure~\ref{fig:clock-vs-sync} maps this landscape.
Each synchronisation method is placed at the clock precision it can serve at a reference averaging time of $\tau = 10{,}000$~s; diagonal lines indicate the matching condition for different averaging times.
Several observations emerge from this comparison.

Classical satellite methods (GPS at ${\sim}5$~ns, TWSTFT at ${\sim}0.3$~ns) are well matched to microwave clocks and commercial oscillators but fall short of optical clock requirements by orders of magnitude.
White Rabbit (${\sim}200$~ps) and ELSTAB (${\sim}50$~ps) serve clocks at the ${\sim}10^{-14}$ to $10^{-15}$ level at $\tau = 10{,}000$~s, adequate for hydrogen masers and commercial oscillators but falling short of caesium fountains (${\sim}10^{-16}$, which would require ${\sim}1$~ps) and all optical clocks.
The quantum methods Q-TWTT (2.46~ps at 20~km~\cite{Hou2019}) and QKD-based timing (4.45~ps at 75~km~\cite{Tang2023}) occupy the 1--10~ps range, matched to clocks at ${\sim}10^{-16}$ for $\tau = 10{,}000$~s, still roughly two orders of magnitude above the best transportable optical clocks (${\sim}10^{-18}$), and three above laboratory systems at $10^{-19}$.
Only the Caldwell et al.\ optical time transfer demonstration (320~as~\cite{Caldwell2023}) has reached the sub-femtosecond regime needed for $10^{-19}$ laboratory clocks, but this system uses a research-grade frequency comb over a folded free-space path and is not deployable.

The practical implication is that time transfer, not clock performance, is now the bottleneck for distributed optical timekeeping.
Coherent optical frequency transfer already operates at the $10^{-19}$ fractional level over continental distances~\cite{Droste2013,Schioppo2022}, but it provides frequency comparison only, not time synchronisation.
Bridging the gap for direct time transfer will require either advancing Q-TWTT from picosecond to sub-femtosecond synchronisation uncertainty, or deploying Caldwell-type comb-based systems operationally, both representing significant but plausible engineering efforts over the coming decade.
In the interim, hybrid architectures that combine coherent optical frequency transfer with quantum or classical time transfer at a coarser level provide a pragmatic approach to comparing next-generation clocks across metropolitan and continental distances.


\section{Use Cases: Where Precise Timing Matters and What Quantum Could Add}\label{sec:use-cases}

This section examines specific domains where precise timing is critical, how timing is addressed today, where vulnerabilities lie, and whether quantum approaches offer a meaningful advantage.
For each domain, we assess whether the quantum value proposition is realistic given current and near-term technology, or whether classical methods will remain sufficient.

\subsection{Financial Markets and Trading}\label{sec:finance}

Financial markets are among the most timing-sensitive civilian applications.
The EU's MiFID~II regulation (RTS~25, effective January 2018) requires high-frequency trading (HFT) firms to synchronise their clocks to within 100~$\mu$s of UTC with 1~$\mu$s timestamp granularity~\cite{Lombardi2016}.
In the United States, the SEC's Consolidated Audit Trail requires exchanges to maintain 100~$\mu$s accuracy, while broker-dealers must achieve 50~ms~\cite{Lombardi2016}.
In practice, HFT firms operate well below these regulatory floors: production systems routinely achieve sub-200~ns synchronisation using PTP with hardware timestamping, because competitive advantage is measured in nanoseconds~\cite{Lombardi2016}.

The timing infrastructure is almost entirely GPS-dependent.
Exchange data centres use GPS-disciplined atomic clocks (rubidium or caesium oscillators) as primary references, distributing time via PTP over the local network.
Lombardi et al.~\cite{Lombardi2016} documented the architecture in detail, showing that NIST-disciplined clocks achieve an uncertainty of $<$15~ns relative to UTC when properly calibrated.

The vulnerability is real and documented.
In 2012, the UK Sentinel project found that GPS signals near the London Stock Exchange were blocked daily for up to ten minutes by a delivery driver's dashboard jammer, disrupting timestamps on financial trades~\cite{Curry2012}.
In January 2016, a software error during the decommissioning of GPS satellite PRN~32 caused 15 of approximately 30 GPS satellites to broadcast incorrect UTC offset parameters, shifting GPS-disciplined clocks by approximately $-13$~$\mu$s~\cite{Lombardi2021}.
Humphreys~\cite{Humphreys2011} showed that GPS spoofing attacks on financial timing infrastructure are technically feasible: by shifting one exchange's clock by several milliseconds relative to another, an attacker could exploit the timing discrepancy for cross-exchange arbitrage.

\textbf{Quantum value assessment.}
The precision requirements of financial markets (sub-$\mu$s) are well within the capability of QKD-based time transfer (${\sim}$4--30~ps).
The value would be in \textit{security}: quantum timing could provide a GPS-independent, spoof-resistant timing reference for exchanging data centres.
However, the practical barrier is significant.
Financial timing operates over local area networks (metres to kilometres), where classical PTP with GPS backup is simple, cheap, and well-understood.
A quantum timing node would need to compete not on precision but on resilience, and the cost of SNSPD-based systems (${\sim}$\texteuro100{,}000 per node) is difficult to justify when the alternative is a redundant GPS receiver (${\sim}$\texteuro10{,}000).
The more realistic near-term path is classical multi-GNSS receivers with holdover oscillators, possibly augmented by fibre-based PTP from a national metrology institute via White Rabbit.
Quantum timing becomes relevant when exchanges are already connected via quantum networks for QKD, in which case timing is a zero-marginal-cost byproduct.

\subsection{Power Grid Synchronisation}\label{sec:power-grid}

Modern power grids rely on phasor measurement units (PMUs) to monitor voltage magnitude and phase angle across the network in real time.
The IEEE C37.118.1 standard specifies a maximum total vector error (TVE) of 1\%, which translates to a timing requirement of approximately 2.6~$\mu$s for a 60~Hz system and 1~$\mu$s recommended resolution~\cite{Lombardi2021}.
More demanding grid applications, such as travelling wave fault location, require 100~ns accuracy~\cite{Lombardi2021}.

PMUs receive timing directly from GPS.
As of 2017, approximately 1{,}800 PMUs were deployed across the North American grid~\cite{Lombardi2021}.
If a PMU receives a spoofed GPS signal, the resulting false phase angle calculation could cause automated control systems to issue incorrect commands.
The consequences are not hypothetical: the 2003 Northeast blackout (affecting 55 million people across 8 US states and 2 Canadian provinces) was not caused by a timing failure, but the NERC post-mortem report noted that the absence of time-synchronised recording severely impeded the forensic investigation, leading to mandatory GPS synchronisation requirements for all grid monitoring equipment~\cite{Lombardi2021}.
A 2019 RTI International study commissioned by NIST estimated that a 30-day GPS outage would cost the US electric power sector \$212--338 million~\cite{Lombardi2021}.

\textbf{Quantum value assessment.}
The precision requirements ($\mu$s level) are easily met by quantum timing, but they are also easily met by classical alternatives (GPS, PTP, White Rabbit).
The security argument is more relevant: GPS spoofing of PMU timing could destabilise grid control, and quantum timing could provide a tamper-resistant alternative for critical substations.
However, the power grid's primary timing need is \textit{availability} (continuous, reliable timing at thousands of distributed nodes), not sub-nanosecond precision.
Deploying quantum timing at 1{,}800+ PMU sites is impractical with current technology.
The realistic path is GPS backup via eLoran, fibre-based PTP from utility communication networks, or improved holdover oscillators.
Quantum timing could serve a niche role at a small number of critical control centres connected to quantum communication networks, but it is not a grid-wide solution.

\subsection{Telecommunications and 5G}\label{sec:telecom}

Telecommunications networks require precise frequency and time synchronisation at every level.
The ITU-T G.811 standard specifies primary reference clock frequency accuracy of $1 \times 10^{-11}$, and the enhanced primary reference time clock (ePRTC, ITU-T G.8272.1) requires $\pm$30~ns accuracy relative to UTC~\cite{Lombardi2021}.
5G New Radio base stations require approximately 1~$\mu$s time synchronisation, with more stringent requirements (${\sim}$tens of nanoseconds) for fronthaul links supporting Coordinated Multi-Point (CoMP) and Massive MIMO features.

Telecom networks are heavily GPS-dependent: a survey of 191 active NTP servers found that 96.9\% were synchronised to GPS~\cite{Lombardi2021}.
CDMA base stations function as GPS repeaters, and the same dependency extends to 4G and 5G infrastructure.
A 30-day GPS outage was estimated to cost the US telecom sector \$5.5--14.2 billion~\cite{Lombardi2021}.

\textbf{Quantum value assessment.}
Telecom timing requirements (ns to $\mu$s) are within quantum capabilities, but the scale of the problem favours classical solutions.
A national 5G network has tens of thousands of base stations, each of which requires timing.
White Rabbit over the existing fibre backhaul network is the natural solution and is already being adopted.
Quantum timing could add value at telecom core nodes or data centres that already host QKD equipment, providing timing as a byproduct.
For the broader base station network, quantum timing is not practical in the near future.

\subsection{Scientific Metrology and Fundamental Physics}\label{sec:metrology}

Scientific applications are the domain where the synchronisation gap identified in Section~\ref{sec:clocks} is most consequential.
Distributed optical clocks synchronised at matching precision would enable relativistic geodesy with centimetre-level geoid resolution~\cite{Lisdat2016,Takamoto2020,Nosske2025}, international comparison of candidate clocks for the planned redefinition of the SI second, and searches for ultralight dark matter and variations of fundamental constants~\cite{Derevianko2014,Filzinger2023}, all requiring correlating clock readings at the $10^{-18}$ level or better across continental distances.

For \textit{frequency comparison}, coherent optical transfer already operates at the $10^{-19}$ fractional level over continental distances~\cite{Droste2013,Schioppo2022}; quantum methods are not needed and cannot compete.
For \textit{direct time synchronisation}, the situation is far less mature.
The best Q-TWTT synchronisation uncertainty is 2.46~ps~\cite{Hou2019}, matched to clocks at the ${\sim}10^{-16}$ level at $\tau = 10{,}000$~s, two to three orders of magnitude above what $10^{-18}$ to $10^{-19}$ clocks require (Section~\ref{sec:sync-gap}).
Q-TWTT's stability (TDEV of 45~fs) is far better than its accuracy, suggesting that systematic error reduction is the path forward, but the current 2--5-fold dispersion cancellation improvement does not by itself close the gap to sub-femtosecond accuracy.

\textbf{Quantum value assessment.}
Scientific metrology is the strongest use case, but current quantum methods fall two to three orders of magnitude short for direct time synchronisation.
The user community (national metrology institutes) is concentrated at a manageable number of fibre-connected sites with a willingness to deploy complex equipment, making this the most natural domain for first demonstrations as Q-TWTT accuracy improves.
VLBI (${\sim}10^{-14}$ stability requirement) is within quantum capabilities but impractical due to the global distribution of stations.
In the longer term, entangled clock networks could enable collective measurements exceeding individual clock precision, but this remains a decade or more away.

\subsection{Military and Defence}\label{sec:military}

Military systems depend heavily on precise timing, and timing requirements vary by orders of magnitude across applications.
Two systems illustrate where synchronisation precision has direct operational consequences.

\textbf{Electronic warfare: TDOA emitter geolocation.}
Time Difference of Arrival (TDOA) is the primary technique for geolocating hostile radar and communications emitters from multiple receiving platforms.
The geolocation error scales directly with synchronisation error: $\Delta x \approx c \cdot \Delta t$, where $c$ is the speed of light and $\Delta t$ is the timing offset between receivers~\cite{ODonoughue2019}.
At 1~ns synchronisation error, the position uncertainty is approximately 30~cm; at 100~ns, approximately 30~m; at 1~$\mu$s, approximately 300~m.
Operational TDOA systems typically require 10--100~m accuracy, corresponding to 30--300~ns synchronisation between platforms separated by tens to hundreds of kilometres~\cite{ODonoughue2019}.
This is well within the capability of military GPS under normal conditions (the Precise Positioning Service specifies $\leq$40~ns at 95\% confidence~\cite{GPSPPS2007}, with modern receivers achieving ${\sim}$5--20~ns in practice), but GPS jamming in contested environments degrades or denies this capability entirely.
GNSS interference rates affecting aviation increased by more than 200\% between 2021 and 2024, with spoofing incidents rising approximately five-fold~\cite{IATA2025safety}, and military operations in conflict zones face substantially worse conditions.

\textbf{Coherent distributed radar.}
Multistatic and distributed radar systems, in which multiple spatially separated transmitters or receivers coherently combine their signals, impose far more stringent timing requirements.
Coherent integration at X-band (${\sim}$10~GHz, $\lambda \approx 3$~cm) requires phase synchronisation better than approximately $\lambda/10$ to maintain useful beamforming gain~\cite{Yang2011mimo}, corresponding to ${\sim}3$~mm path-length agreement and timing synchronisation of approximately 10~ps between nodes.
This is beyond the capability of GPS (${\sim}$5--40~ns) by three orders of magnitude, placing coherent distributed radar in the regime where only fibre-based methods (White Rabbit, ELSTAB, or quantum approaches) can provide adequate timing.
For fixed installations connected by fibre, White Rabbit (${\sim}$200~ps) approaches but does not meet the 10~ps requirement; Q-TWTT, with a synchronisation uncertainty of 2.46~ps, would be sufficient but has not been demonstrated in a military context.

\textbf{Quantum value assessment.}
The core vulnerability in military timing is GPS dependence in adversarial environments.
US Executive Order 13905 (February 2020) mandated development of GPS-independent timing sources, classifying PNT services as a ``largely invisible utility''~\cite{Lombardi2021}.
For TDOA-class applications (30--300~ns requirement), the precision of quantum timing (${\sim}$4--30~ps) is massive overkill; the value lies entirely in \textit{security and resilience}, providing a GPS-independent, tamper-resistant timing reference over military fibre networks via QKD-based time transfer.
For coherent distributed radar (${\sim}$10~ps requirement), quantum timing is one of the few approaches with sufficient precision, though this applies only to fixed, fibre-connected installations, not to mobile or airborne platforms.
The practical barriers are significant: QKD-based systems require dedicated fibre links and, for the highest precision, cryogenic detectors, limiting deployment to permanent military infrastructure rather than field-deployed forces.
Satellite-based quantum timing (building on the Micius demonstration~\cite{Dai2020}) could serve mobile platforms, but current precision (30--60~ps) and availability (weather-dependent, brief satellite passes) are inferior to military GPS.

\subsection{Quantum Networks as Timing Infrastructure}\label{sec:qnet-timing}

The most practical near-term path to quantum timing deployment is not a standalone application but rather a byproduct of deploying quantum networks for other purposes.
Multiple countries and the EU (via the EuroQCI initiative) are investing in quantum communication infrastructure primarily for QKD.
These networks inherently distribute time-correlated photon pairs, and extracting timing information requires only software, not additional hardware~\cite{Spiess2023,Tang2023}.
Tang et al.~\cite{Tang2023} demonstrated this in a 75~km entanglement distribution network, achieving 4.45~ps synchronisation uncertainty, precision that exceeds the requirements of telecommunications, power grids, and financial systems.
Rather than competing head-to-head with classical timing, quantum timing enters as an added capability of the quantum communication infrastructure being built for security reasons.

\subsection{Summary}\label{sec:use-case-summary}

Table~\ref{tab:use-case-summary} assesses the quantum value proposition across application domains.

\begin{table}[ht]
\centering
\caption{Assessment of the quantum timing value proposition by application domain.}
\label{tab:use-case-summary}
\small
\begin{tabularx}{\textwidth}{>{\raggedright\arraybackslash}p{2.2cm} >{\raggedright\arraybackslash}p{2.2cm} >{\raggedright\arraybackslash}p{2cm} >{\raggedright\arraybackslash}p{2.2cm} X}
\hline
\textbf{Domain} & \textbf{Required precision} & \textbf{Current solution} & \textbf{Quantum advantage} & \textbf{Assessment} \\
\hline
Financial markets & $<$1~$\mu$s (reg.); $<$200~ns (practice) & GPS + PTP & Security & Niche: only viable if QKD network already exists \\[4pt]
Power grid & 1--3~$\mu$s (PMU) & GPS & Security & Impractical at grid scale (1800+ PMU sites) \\[4pt]
Telecom / 5G & 30~ns -- 1~$\mu$s & GPS + PTP / White Rabbit & Security & Viable at core nodes only \\[4pt]
Scientific metrology & $10^{-19}$ fractional (freq.); sub-fs (time) & Coherent optical (freq.); none adequate (time) & Dispersion cancellation (time sync) & Strongest case, but Q-TWTT 2--3 orders from time sync target \\[4pt]
Military: TDOA/EW & 30--300~ns & Military GPS & Security (GPS-denied) & Precision overkill; value is resilience \\[4pt]
Military: coherent radar & ${\sim}$10~ps & GPS insufficient & Precision + security & Viable for fixed fibre-connected sites only \\[4pt]
Quantum networks & ${\sim}$4--30~ps & N/A (new) & Zero-cost byproduct & Most practical near-term path \\
\hline
\end{tabularx}
\end{table}

In summary, quantum time synchronisation is not needed for its \textit{precision} in most applications, and even in the one domain where precision matters most, scientific metrology, current quantum methods fall two to three orders of magnitude short of what next-generation optical clocks require for direct time synchronisation (Section~\ref{sec:sync-gap}).
The \textit{security} properties of quantum timing are valuable in adversarial environments, and \textit{integration} with quantum networks provides the most practical deployment path.
The domains where quantum timing is most likely to see early adoption are military/defence (for security), quantum communication networks (as a zero-cost byproduct), and scientific metrology (as Q-TWTT accuracy improves toward the sub-femtosecond level needed for optical clock comparisons).


\section{Conclusion}
\label{sec:conclusion}

The theoretical advantages of quantum time synchronisation, such as Heisenberg-limited scaling and nonlocal dispersion cancellation, are real but bounded: decoherence reduces Heisenberg scaling to constant-factor improvements~\cite{DemkowiczDobrzanski2012}, and dispersion cancellation provides a 2--5-fold gain rather than complete elimination.
Quantum time synchronisation does not outperform deployed classical systems such as White Rabbit or ELSTAB in raw timing precision, although Q-TWTT achieves comparable performance over metropolitan distances~\cite{Hou2019,Hong2024,Shi2024}.

The most significant finding is the widening gap between clock performance and time-transfer capability.
Optical clocks have reached systematic uncertainties below $10^{-18}$, yet the best demonstrated synchronisation uncertainty is 2.46~ps~\cite{Hou2019}, matched to clocks at the ${\sim}10^{-16}$ level, a shortfall of two to three orders of magnitude.
A $10^{-19}$ clock synchronised via a picosecond-class link operates, for all practical purposes, as a $10^{-16}$ clock.
Closing this gap is the most critical open challenge in the field, and the area where Q-TWTT's dispersion cancellation provides the strongest fundamental motivation.

For the vast majority of timing-critical applications, classical methods are sufficient and quantum timing adds value primarily through physical-layer security~\cite{LamasLinares2018} or as a zero-cost byproduct of quantum communication networks~\cite{Tang2023,Spiess2023}.
The most realistic near-term deployment model is a hybrid architecture where quantum timing provides security monitoring and redundancy alongside classical systems.
Looking ahead, the key developments will be in reducing Q-TWTT systematic uncertainties toward the sub-femtosecond level, improved detector technology (higher-$T_c$ SNSPDs), and cascaded relay architectures for range extension~\cite{Hong2026cascaded}.

\printbibliography

\end{document}